\documentclass[11pt]{article}

\usepackage{latexsym,amsmath,amssymb,amscd,eucal}
\usepackage{xypic}
\usepackage{enumerate}

\def\harr#1#2{\smash{\mathop{\hbox to .3in{\rightarrowfill}}
 \limits^{\scriptstyle#1}_{\scriptstyle#2}}}

\def\appendix#1{\addtocounter{section}{1}\setcounter{equation}{0}
\renewcommand{\thesection}{\Alph{section}}
\section*{Appendix \thesection\protect\indent \parbox[t]{11.715cm} {#1}}
\addcontentsline{toc}{section}{Appendix \thesection\ \ \ #1} }
\newcommand{\eq}{\begin{equation}}
\newcommand{\eqend}{\end{equation}}

\newbox\ncintdbox \newbox\ncinttbox
\setbox0=\hbox{$-$} \setbox2=\hbox{$\displaystyle\int$}
\setbox\ncintdbox=\hbox{\rlap{\hbox to
\wd2{\hskip-.125em\box2\relax \hfil}}\box0\kern.1em}
\setbox0=\hbox{$\vcenter{\hrule width 4pt}$}
\setbox2=\hbox{$\textstyle\int$}
\setbox\ncinttbox=\hbox{\rlap{\hbox to
\wd2{\hskip-.175em\box2\relax \hfil}}\box0\kern.1em}

\newcommand{\Tr}[1]{\:{\rm Tr}\,#1}

\def\be{\begin{equation}}
\def\ee{\end{equation}}
\def\bea{\begin{eqnarray}}
\def\eea{\end{eqnarray}}
\def\bd{\begin{displaymath}}
\def\ed{\end{displaymath}}

\DeclareFontFamily{U}{rsf}{}
\DeclareFontShape{U}{rsf}{m}{n}{
  <5> <6> rsfs5 <7> <8> <9> rsfs7 <10-> rsfs10}{}
\DeclareMathAlphabet\Scr{U}{rsf}{m}{n}

\makeatletter
\newdimen\normalarrayskip              
\newdimen\minarrayskip                 
\normalarrayskip\baselineskip
\minarrayskip\jot
\newif\ifold             \oldtrue            
\def\arraymode{\ifold\relax\else\displaystyle\fi} 
\def\@arrayskip{\ifold\baselineskip\z@\lineskip\z@
     \else
     \baselineskip\minarrayskip\lineskip2\minarrayskip\fi}
\def\@arrayclassz{\ifcase \@lastchclass \@acolampacol \or
\@ampacol \or \or \or \@addamp \or
   \@acolampacol \or \@firstampfalse \@acol \fi
\edef\@preamble{\@preamble
  \ifcase \@chnum
     \hfil$\relax\arraymode\@sharp$\hfil
     \or $\relax\arraymode\@sharp$\hfil
     \or \hfil$\relax\arraymode\@sharp$\fi}}
\def\@array[#1]#2{\setbox\@arstrutbox=\hbox{\vrule
     height\arraystretch \ht\strutbox
     depth\arraystretch \dp\strutbox
     width\z@}\@mkpream{#2}\edef\@preamble{\halign \noexpand\@halignto
\bgroup \tabskip\z@ \@arstrut \@preamble \tabskip\z@ \cr}%
\let\@startpbox\@@startpbox \let\@endpbox\@@endpbox
  \if #1t\vtop \else \if#1b\vbox \else \vcenter \fi\fi
  \bgroup \let\par\relax
  \let\@sharp##\let\protect\relax
  \@arrayskip\@preamble}
\makeatother

\newcommand{\beq}{\begin{eqnarray}}
\newcommand{\eeq}{\end{eqnarray}}

\def\appendix#1{\addtocounter{section}{1}\setcounter{equation}{0}
\renewcommand{\thesection}{\Alph{section}}
\section*{Appendix \thesection. #1}
\addcontentsline{toc}{section}{Appendix \thesection\ \ \ #1} }

\numberwithin{equation}{section}

\begin{document}


\vspace{.1in}

\begin{center}

{\Large\bf PARTITION FUNCTIONS OF THREE-DIMENSIONAL QUANTUM GRAVITY AND THE BLACK HOLE ENTROPY}

\end{center}
\vspace{0.1in}
\begin{center}
{\large
A. A. Bytsenko $^{(a)}$
\footnote{abyts@uel.br}},
and
\, M. E. X. Guimar\~aes $^{(b)}$ \footnote{emilia@if.uff.br}

\vspace{7mm}

$^{(a)}$ {\it
Departamento de F\'{\i}sica, Universidade Estadual de
Londrina\\
Caixa Postal 6001, Londrina-Paran\'a, Brazil}
\vspace{5mm}\\
$^{(b)}$ {\it Instituto de F\'{\i}sica,
Universidade Federal Fluminense,\\
Av. Gal. Milton Tavares de Souza s/n, Niter\'oi-RJ, Brazil}
\vspace{5mm}\\
\end{center}
\vspace{0.1in}
\begin{center}
{\bf Abstract}
\end{center}
We analyze aspects of the holographic principle relevant to the quantum gravity partition functions in Euclidean sector of AdS$_3$. The sum of known contributions to the partitions functions can be presented exactly, including corrections,  
in the form where the Patterson-Selberg zeta function involves.

\vfill



\newpage


\section{Introduction}

Recently quantum gravity partition functions in the locally three-dimensional Anti-de Sitter (AdS$_3$) space-times have been analyzed in detail \cite{Maloney,Giombi}. In general such an analisis relies on the relation of the relevant product of determinants to
Ray-Singer and Reidemeister torsion. For an
irreducible flat connection on a compact manifold without
boundary, the torsion is simply a number, but 
for a manifold with boundary (such manifolds were considered in \cite{Maloney,Giombi}), the torsion must be understood as the measure on a certain moduli space associated with
boundary. We suspect that computation of the one-loop correction using its interpretation via torsion is rather complicate. Nevertheless, the one-loop correction we may
study by expressing the determinants in terms of the appropriate heat kernels, which can be evaluated by a images method (also one can use the Selberg trace formula applied to the tensor Laplacian). It has been shown that
the symmetry group of AdS$_3$ gravity (with appropriate boundary conditions) is generated by the Virasoro algebra \cite{Brown}, and the one-loop partition function of gravity in AdS$_3$ does endeed the partition function of a conformal field theory (CFT$_2$) in two dimensions \cite{Giombi}.  

Now let us ask to what extent there may be AdS$_3$/CFT$_2$ correspondence, if it is completely correct. It is known that a simple geometrical structure of three-dimensional gravity (and black holes) allows exact computations since its Euclidean counterpart is locally isomorphic to the constant curvature hyperbolic space. 
Since the correspondence between quantum gravity partition function in AdS$_3$ and contribution to the partition function of CFT$_2$, we respect also correspondence between spectral functions related to Euclidean AdS$_3$ 
and modular-like functions (a Poincar\'e series) associated with a complex Riemann suirface. We assume that this correspondence occur when complex variables of spectral functions take values on a Riemann surface (i.e. on the conformal boundary of AdS$_3$).

Thus, the main purpose of this work is to analyze a correspondence between the partition functions (and spectral functions) of three-dimensional quantum gravity and CFT in two-dimensions and provide the holographic principle relevant to this coprrespondence.

\section{Expository remarks on partition functions}

In this section there are no original results; we discuss partition functions for three-dimensional gravity and ${\mathcal N=1}$ supergravity mostly followed \cite{Maloney}.

{\bf Quantum graity in three dimensions.}
It has been shown that the contribution to the
partition function of pure gravity in a space-time asymptotic to
$AdS_3$ comes from smooth geometries $M =\Gamma\backslash AdS_3$, where $\Gamma$ is a discrete subgroup of $SO(3,1)$. 
To be more precise, it comes from geometries $M_{c,d}$\, (see for detail \cite{Maloney}), where $c$ and $d$ are a pair of relatively prime integers, $c\geq 0$, and
a pair $(c,d)$ identified with $(-c,-d)$. The manifolds
$M_{c,d}$ are all diffeomorphic to each other, and therefore the contribution ${\bf Z}_{c,d}(\tau)$ to the partition function can be expressed in terms of any one of them, say ${\bf Z}_{0,1}(\tau)$, by a modular transformation. It gives the following formula: 
$
{\bf Z}_{c,d}(\tau)={\bf Z}_{0,1}((a\tau+b)/(c\tau+d)),
$
where 
\begin{equation}
{\bf Z}_{0,1}(\tau)= |q \bar q|^{-k} \prod_{m=2}^\infty|1-q^m|^{-2}\,.
\label{PF}
\end{equation} 
The modulos of a Riemann surface $\Sigma$ of genus one (the conformal boundary of AdS$_3$) is defined up to $\gamma \tau = (a\tau+ b)/(c\tau + d)$ with 
$\gamma = \left(a\ b \atop c\ d\right) \in
SL(2,{\mathbb Z})\,.
$
The partition function as the sum of known contributions 
of states of left- and right-moving modes in the conformal field theory takes the form \cite{Maloney}
\begin{equation}
{\bf Z}(\tau)=\sum_{c,d}{\bf Z}_{c,d}(\tau)=
\sum_{c,d}{\bf Z}_{0,1}((a\tau+b)/(c\tau+d))\,.
\label{Z02}
\end{equation}

{\bf NS sector of \,${\mathcal N}$=1 supergravity.} We will consider only the basic case of ${\mathcal N}$ =1 supergravity. Thus the symmetry group $SL(2,\mathbb R)\times SL(2,\mathbb R)$
of $AdS_3$ is replaced by $OSp(1|2)\times OSp(1|2)$, where
$OSp(1|2)$ is a supergroup whose bosonic part is
$Sp(2,\mathbb R)=SL(2,\mathbb R)$. The boundary CFT has $(1,1)$ supersymmetry (${\mathcal N}=1$ supersymmetry for both left- and right-movers). There are a few closely related choices of possible partition function,  
$
{\rm Tr}\,\exp(-\beta H-i\theta J)
$
or
$
{\rm Tr}\,(-1)^F\exp(-\beta H-i\theta J)\,.
$
This trace  could be computed in either the Neveu-Schwarz (NS) or the Ramond (R) sector. One can compute these partition functions  by summing over three-manifolds $M$ that are locally $AdS_3$ and whose conformal boundary is a Riemann surface $\Sigma$ of genus one. In addition, the four possible
partition functions associated with NS or R sectors (with or without an insertion of $(-1)^F$) correspond to the four spin structures on $\Sigma$. An element $g$ of $G=SL(2, {\mathbb R})$ acts on spin structure by
\begin{equation}
g
\left[ \begin{array}{c} \mu \\ \nu \end{array} \right]
\longrightarrow
\left[ \begin{array}{cc} a & b \\ c & d \end{array} \right]
\left[ \begin{array}{c} \mu \\ \nu \end{array} \right],
\,\,\,\,\,\, \mu, \nu \in (1/2){\mathbb Z}/{\mathbb Z}\,,
\label{structure}
\end{equation}
where the four spin structures on the two-torus $\Sigma$ is represented by the column vector in (\ref{structure}), and $\mu, \nu$ taking the values (1/2) for antiperiodic (NS) boundary conditions and 0 for periodic (R) ones.

Taking into account the choice of the spin structure on $\Sigma$, one can sum over choices of $M$ such that the given spin structure on $\Sigma$ does extend over $M$. The NS spin structure on $\Sigma$ is compatible with $M_{0,1}$, and therefore $M_{0,1}$ contributes to traces in the NS sector, not the R sector. The partition function of left- and right-moving exitations is $F(q, \overline{q}) = {\rm Tr}_{\rm NS}\exp(-\beta H - i\theta J)$, and the contribution to $F(q, \overline{q})$ associated with $M_{0,1}$ becomes \cite{Maloney}:
\begin{equation}
F_{0,1}= F_{0,1}^{(\rm ground)}\cdot {\widehat F}_{0,1}(\tau)
\equiv
\left|q^{-k^*/2}\right|^2\cdot
\left|\prod_{n=2}^\infty{1+q^{n-1/2}\over
1-q^n}\right|^2\,, 
\label{F01}
\end{equation}  
where the contribution $F_{0,1}^{(\rm ground)}\equiv \left|q^{-k^*/2}\right|^2$ is related to the ground state energy. 
The complete function $F(\tau)$ can be computed by summing $F_{0,1}$ over modular images with $c+d$ odd. It corresponds to spin structure with $\mu=\nu = 1/2$.
Thus,
\begin{equation}
F {\left[ \begin{array}{c} \frac{1}{2} \\ \frac{1}{2} \end{array} \right]}
(\tau)=\sum_{c,d|c+d\,\,
{\rm odd}}{\widehat F}_{0,1}((a\tau+b)/(c\tau+d)).
\label{F}
\end{equation}
Let us also analyze partition functions with other spin structures. If we let $\mu=0$,
$\nu=1/2$, then we get $ G(q, \overline{q})=\Tr_{\rm NS}\,(-1)^F\exp(-\beta
H-i\theta J)$, and the contribution of $M_{0,1}$ to this partition function is obtained by reversing the sign of all fermionic contributions in (\ref{F01}): 
\begin{equation}
G_{0,1}(\tau)= G_{0,1}^{(\rm ground)}\cdot 
{\widehat G}_{0,1}(\tau)
\equiv
\left|q^{-k^*/2}\right|^2\cdot
\left|\prod_{n=2}^\infty{1-q^{n-1/2}\over
1-q^n}\right|^2.
\label{G01}
\end{equation} 
For the spin structure with $\mu=0$ and $\nu=1/2$ one has
\begin{equation}
G{\left[ \begin{array}{c} 0 \\ \frac{1}{2} \end{array}
\right]    }
(\tau)=\sum_{c,d|d\,\,{\rm odd}}
{\widehat G}_{0,1}((a\tau+b)/(c\tau+d))\,.
\label{G}
\end{equation} 
Note that the summand in (\ref{F}) and (\ref{G}) does not depend on the choice $a, b$ \cite{Maloney}. 
A modular transformation
$\tau\to\tau+1$ exchanges pair $(\mu,\nu)=(0,1/2)$ with
$(\mu,\nu)=(1/2,1/2)$; in particular,
$
F(\tau)=G(\tau+1)=F(\tau+2)\,.
$

{\bf R sector of ${\mathcal N}$=1 supergravity.}
One can compute the Ramond partition function $K={\rm Tr}_{\rm R}\exp(-\beta H-i\theta J)$ for $\mu=1/2$, $\nu=0$, so
$
K(\tau)=G(-1/\tau).
$ 
This completes characterization list of three of the four partition functions. In supersymmetric theory with discrete spectrum, the fourth partition
function $I=\Tr_{\rm R}(-1)^F\exp(-\beta H-i\theta J)$ is an
integer, independent of $\beta$ and $\theta$ (it can be
interpreted as the index of a supersymmetry generator). 
This function has to be computed using the odd spin structure, the one with $\mu=\nu=0$. 
Typically in three-dimensional gravity the partition function $I$ vanishes, since the odd spin structure does not extend over any three-manifold with boundary $\Sigma$.

\section{Holomorphic factorization}

\subsection{Spectral functions of hyperbolic geometry}
The Euclidean sector of $AdS_3$ has an orbifold description ${\bf H}_\Gamma = \Gamma\backslash H^3$. 
The complex unimodular group $G=SL(2, {\mathbb C})$
act on real hyperbolic three-space $H^3$ in a standard way, namely for $(x,y,z)\in H^3$ and $g\in G$, 
$g\cdot(x,y,z)= (u,v,w)\in H^3$. Thus for $r=x+iy$,\,
$g= \left[ \begin{array}{cc} a & b \\ c & d \end{array} \right]$,
\begin{equation}
u+iv = \frac{(ar+b)\overline{(cr+d)}+ a\overline{c}z^2}
{|cr+d|^2 + |c|^2z^2},\,\,\,\,\,\,\,
w = \frac{z}
{|cr+d|^2 + |c|^2z^2}\,.
\end{equation}
Here the bar denotes the complex conjugation. Let $\Gamma \in G$ be the discrete group of $G$.
Define the discrete group $\Gamma$, 
\begin{eqnarray}
\Gamma & = & \{{\rm diag}(e^{2n\pi ({\rm Im}\,\tau + i{\rm Re}\,\tau)},\,\,  e^{-2n\pi ({\rm Im}\,\tau + i{\rm Re}\,\tau)}):
n\in {\mathbb Z}\}
= \{{\frak g}^n:\, n\in {\mathbb Z}\}\,,
\nonumber \\
{\frak g} & = &
{\rm diag}(e^{2\pi ({\rm Im}\,\tau + i{\rm Re}\,\tau)},\,\,  e^{-2\pi ({\rm Im}\,\tau + i{\rm Re}\,\tau)})\,.
\end{eqnarray}
One can define the Selberg zeta function for the group 
$\Gamma = \{{\frak g}^n : n \in {\mathbb Z}\}$ generated by a single hyperbolic element of the form ${\frak g} = {\rm diag}(e^z, e^{-z})$, where $z=a+ib$ for $a,b >0$. In fact we will take
$a = 2\pi {\rm Im}\,\tau$, $b= 2\pi {\rm Re}\,\tau$. For the standard action of $SL(2, {\mathbb C})$ on $H^3$ one has
\begin{equation}
{\frak g}
\left[ \begin{array}{c} x \\ y\\ z \end{array} \right]
=
\left[\begin{array}{ccc} e^{a} & 0 & 0\\ 0 & e^{a} & 0\\ 0
& 0 & \,\,e^{a} \end{array} \right]
\left[\begin{array}{ccc} \cos(b) & -\sin (b) & 0\\ 
\sin (b) & \,\,\,\,\cos (b) & 0
\\ 0 & 0 & 1 \end{array} \right]
\left[\begin{array}{c} x \\ y\\ z \end{array} \right]
\,.
\end{equation}
Therefore ${\frak g}$ is the composition of a rotation in ${\mathbb R}^2$ with complex eigenvalues $\exp (\pm ib)$ and a dilation $\exp (a)$.
The following zeta function can be attached to ${\bf H}_\Gamma$ (see for detail \cite{Bytsenko1}):
\begin{equation}
Z_\Gamma(s) :=\prod_{\stackrel{k_1,k_2\geq
0}{k_1,k_2\in\mathbb{Z}}}^\infty[1-(e^{i\theta})^{k_1}(e^{-i\theta})^{k_2}e^{-(k_1+k_2+s)\ell}]\,.
\label{zeta}
\end{equation}
We should emphasize that ${\bf H}_{\Gamma}$ is also the geometry of a Euclidean three-dimensional black hole. In (\ref{zeta}) $\ell = 2\pi {\rm Im}\, \tau = 2\pi r_{+},\, 
\theta = 2\pi {\rm Re}\, \tau = 2\pi |r_{-}|$, where
$r_{+}> 0$,\, and $r_{-} \in i{\mathbb R}\, (i^2=-1)$ are the outer and inner horizons of a black hole. 
$Z_\Gamma(s)$ is an entire function of $s$, whose zeros are precisely the complex numbers
$
\zeta_{n,k_{1},k_{2}} = -\left(k_{1}+k_{2}\right)+i\left(k_{1}-
k_{2}\right)\theta/\ell+ 2\pi  in/\ell$\,, 
($n \in {\mathbb Z}$), and whose logarithm for ${\rm Re} s> 0$ is given by \cite{Bytsenko1}
\begin{equation}
{\rm log}\, Z_\Gamma (s) =
-\frac{1}{4}\sum_{n=1}^{\infty}\frac{e^{-n\ell(s-1)}}
{n[\sinh^2\left(\frac{\ell n}{2}\right)
+\sin^2\left(\frac{\theta n}{2}\right)]}
=
-\frac{1}{4}\sum_{n=1}^{\infty}\frac{e^{-2n\pi{\rm Im}\,\tau(s-1)}}
{n|\sin(n\pi \tau)|^2}\,.
\label{zeta2}
\end{equation}

\subsection{Quantum corrections}
For three-dimensinal gravity in real hyperbolic space the one-loop partition function, as a product of holomorphic and antiholomorphic functions, has been calculated in \cite{Giombi}, and the result is: 
\begin{equation}
{\widehat {\bf Z}}(\tau)= \left[\prod_{m =2}^{\infty}[(1-q^m)_{\rm hol}\prod_{m=2}^{\infty}(1-\overline{q}^m)_{\rm antihol}
\right]^{-1} =
\prod_{m =2}^{\infty}|1-q^m|^{-2}.
\end{equation}
We may evaluate ${\rm log}\,{\widehat {\bf Z}}(\tau)$ to get
the formula
\begin{eqnarray}
{\rm log}\, {\widehat {\bf Z}}(\tau) & = & 
\sum_{n=1}^{\infty}\frac{e^{-2n\pi {\rm Im}\,\tau}
\cos (4n\pi {\rm Re}\,\tau) - e^{-4n\pi {\rm Im}\,\tau}
\cos (2n\pi {\rm Re}\,\tau)}
{2n[\sinh^2\left(n\pi {\rm Im}\tau\right)
+\sin^2\left(n\pi{\rm Re}\tau\right)]}
\nonumber \\
& = &
{\rm log}\left\{
\left[\frac{Z_\Gamma(3- i\frac{{\rm Re}\tau}{{\rm Im}\tau})}
{Z_\Gamma(2- 2i\frac{{\rm Re}\tau}{{\rm Im}\tau})}\right]_{\rm hol}
\cdot \left[\frac{
Z_\Gamma(3 + i\frac{{\rm Re}\tau}{{\rm Im}\tau})}
{Z_\Gamma(2 + 2i\frac{{\rm Re}\tau}{{\rm Im}\tau})}\right]_{\rm antihol}\right\}.
\label{final0}
\end{eqnarray}
The expression (\ref{final0}) is invariant under  the transformation $\tau \rightarrow \tau + 1$. 
We would like to comment on the sum over geometries.
The partition function, including the contribution from the Brown-Henneaux excitations, have the form
\begin{eqnarray} 
\!\!\!\!
{{\bf Z}}(\tau) = \sum_{c,d} 
{\bf Z}_{c, d}(\gamma\tau) & = &
\sum_{c,d}
\left|q^{-k}
\prod_{n=2}^{\infty}(1-q^{n})^{-1}\right|_{\gamma}^{2}
\nonumber \\
& = & 
\sum_{c,d} \left\{
|q\overline{q}|^{-k}\cdot
\left[
\frac{Z_\Gamma (3- it)}
{Z_\Gamma (2- 2it)}
\right]_{{\rm hol}}
\cdot\left[\frac{Z_\Gamma(3 + it)}
{Z_\Gamma(2 + 2it)}
\right]_{{\rm antihol}}
\right\}_{\gamma}.
\label{summand}
\end{eqnarray}
Here $t = {\rm Re}\,\tau/{\rm Im}\,\tau$ and $| ... |_{\gamma}$ denote the transform of an expression $| ... |$ by $\gamma$.
As we have seen (Eq. (\ref{final0})) the quantum contribution to ${\widehat {\bf Z}}(\tau)$ is invariant under $\tau\to\tau+1$,  and therefore the summand in (\ref{summand}) is independent of the choice of $a$ and $b$ in $\gamma$.
Note that sum over $c$ and $d$ in (\ref{summand}) should be thought of as a sum over the coset $PSL(2,\mathbb Z)/{\mathbb Z} \equiv (SL(2,\mathbb Z)/\{\pm 1\})/\mathbb Z$.  
Since $ad-bc =1$, we get $(a \tau + b)/(c \tau + d) = a/c\,\, -1/c(c\tau +d)$, one may show that
\begin{equation}
{\rm Im}\,(\gamma\tau)=\frac{{\rm Im}\,\tau}
{|c\tau+d|^2}\,,\,\,\,\,\,\,\,
{\rm Re}\,(\gamma\tau)= \frac{a}{c}-\frac{c{\rm Re}\,\tau+d}{c|c\tau+d|^2}\,.
\end{equation}
The final sums (\ref{summand}) are divergent. These kind of divergences also have been encountered in similar sums associated with three-dimensional gravity
\cite{Dijkgraaf,Kleban,Manschot}.

We will now analyze the partiton function in more details.
Our goal here will be to repeat the analysis of Sect. 2 by using the spectral functions representation.
Useful partition functions are presented in Table \ref{Table1},
where $\xi \equiv (2{\rm Im}\,\tau)^{-1}$.
\begin{table}\label{Table1}
\begin{center}
\begin{tabular}
{l l}
Table {}  \ref{Table1}. List of partition functions
\\
\\
\hline
\\
$ {\rm log}\prod_{m=2}^{\infty}(1-q^m) = -\sum_{n=1}^{\infty}
\frac{q^{2n}}{(1-q^n)}$ & \,\,\,\,\, $ \prod_{m=2}^{\infty}(1-q^m)=
\frac{Z_\Gamma(2-2it)}
{Z_\Gamma(3-it)}$ \\
\\
$ {\rm log}\prod_{m=2}^{\infty}(1-\overline{q}^m) = -\sum_{n=1}^{\infty}
\frac{\overline{q}^{2n}}{(1-\overline{q}^n)}$ & \,\,\,\,\, $ \prod_{m=2}^{\infty}(1-\overline{q}^m)=
\frac{Z_\Gamma(2+2it)}
{Z_\Gamma(3+it)}$ \\
\\
$ {\rm log}\prod_{m=2}^{\infty}(1+q^m) = -\sum_{n=1}^{\infty}
\frac{(-1)^n q^{2n}}{n(1-q^n)}$ & \,\,\,\,\, $ \prod_{m=2}^{\infty}(1+q^m) =
\frac{Z_\Gamma(2-2it-i\xi)}
{Z_\Gamma(3-it-i\xi)}$
\\
\\
$ {\rm log}\prod_{m=2}^{\infty}(1+\overline{q}^m) = -\sum_{n=1}^{\infty}
\frac{(-1)^n \overline{q}^{2n}}{n(1-\overline{q}^n)}$ & \,\,\,\,\, $ \prod_{m=2}^{\infty}(1+\overline{q}^m) =
\frac{Z_\Gamma(2+2it-i\xi)}
{Z_\Gamma(3+it-i\xi)}$
\\
\\
$ {\rm log}\prod_{m=2}^{\infty}(1-q^{m-\frac{1}{2}}) = \sum_{n=1}^{\infty}
\frac{q^{\frac{3n}{2}}}{n(1-q^n)}$ & \,\,\,\,\, $ \prod_{m=2}^{\infty}(1-q^{m-\frac{1}{2}}) =
\frac{Z_\Gamma(\frac{5}{2}-\frac{1}{2}it)}
{Z_\Gamma(\frac{3}{2}-\frac{3}{2}it)}$
\\
\\
$ {\rm log}\prod_{m=2}^{\infty}(1-\overline{q}^{m-\frac{1}{2}}) = \sum_{n=1}^{\infty}
\frac{\overline{q}^{\frac{3n}{2}}}{n(1-\overline{q}^n)}$ & \,\,\,\,\, $ \prod_{m=2}^{\infty}(1-\overline{q}^{m-\frac{1}{2}}) =
\frac{Z_\Gamma(\frac{5}{2}+\frac{1}{2}it)}
{Z_\Gamma(\frac{3}{2}+\frac{3}{2}it)}$
\\
\\
$ {\rm log}\prod_{m=2}^{\infty}(1+q^{m-\frac{1}{2}}) = \sum_{n=1}^{\infty}
\frac{(-1)^n q^{\frac{3n}{2}}}{n(1-q^n)}$ &\,\,\,\,\, $ \prod_{m=2}^{\infty}(1+ q^{m-\frac{1}{2}}) =
\frac{Z_\Gamma(\frac{3}{2}-\frac{3}{2}it-i\xi)}
{Z_\Gamma(\frac{5}{2}-\frac{1}{2}it-i\xi)}$
\\
\\
$ {\rm log}\prod_{m=2}^{\infty}(1+\overline{q}^{m-\frac{1}{2}}) = \sum_{n=1}^{\infty}
\frac{(-1)^n \overline{q}^{\frac{3n}{2}}}{n(1-\overline{q}^n)}$ & \,\,\,\,\, $ \prod_{m=2}^{\infty}(1+ \overline{q}^{m-\frac{1}{2}}) =
\frac{Z_\Gamma(\frac{3}{2}+\frac{3}{2}it-i\xi)}
{Z_\Gamma(\frac{5}{2}+\frac{1}{2}it-i\xi)}$
\\
\\
\hline
\end{tabular}
\end{center}
\end{table}
Since
\begin{eqnarray}
\!\!\!\!\!\!\!\!\!\!
{\rm log}\prod_{m=2}^{\infty}|1+q^{m-\frac{1}{2}}|^2
& = & - \sum_{n=1}^{\infty}\frac{(-1)^n|q|^{-n}[q^{\frac{3n}{2}} +
\overline{q}^{\frac{3n}{2}}-|q|^{2n}(q^{\frac{n}{2}}+
\overline{q}^{\frac{n}{2}})]}{4n|\sin (n\pi \tau)|^2}
\nonumber \\
& = &
- \sum_{n=1}^{\infty}\frac{(-1)^n[e^{-n\pi {\rm Im}\,\tau}
\cos (3n\pi {\rm Re}\,\tau) -e^{-3n\pi {\rm Im}\,\tau}
\cos (n\pi {\rm Re}\,\tau)]}
{2n |\sin (n\pi \tau)|^2}\,,
\end{eqnarray}
then, using the result from Table {\ref{Table1}}, we get
\begin{eqnarray}
{\widehat F}_{0,1}(\tau) 
& = & 
\left[\frac{\prod_{n=2}^\infty (1+q^{n-1/2})} 
{\prod_{n=2}^\infty (1-q^{n})}\right]_{\rm hol}
\cdot
\left[\frac{\prod_{n=2}^\infty (1+\overline{q}^{n-1/2})}
{\prod_{n=2}^\infty (1-\overline{q}^{n})} \right]_{\rm antihol}
\nonumber \\
& = &
\left[\frac{Z_\Gamma(\frac{3}{2}-\frac{3}{2}it-i\xi)
Z_\Gamma(3-it)}
{Z_\Gamma(\frac{5}{2}-\frac{1}{2}it-i\xi)Z_\Gamma(2-2it)}
\right]_{\rm hol}
\left[\frac{Z_\Gamma(\frac{3}{2}+\frac{3}{2}it-i\xi)
Z_\Gamma(3+it)}{Z_\Gamma(\frac{5}{2}+\frac{1}{2}it-i\xi)
Z_\Gamma(2+2it)}\right]_{\rm antihol}\!\!\!. 
\end{eqnarray}
The complete function $F(\tau)$ becomes
\begin{eqnarray}
F{\left[ \begin{array}{c} \frac{1}{2} \\ \frac{1}{2} \end{array} \right]}
(\tau) & = & 
\sum_{c,d|c+d\,\,
{\rm odd}} F_{0,1}^{(\rm ground)}(\gamma\cdot \tau)
{\widehat F}_{0,1}(\gamma \cdot\tau)
\nonumber \\
&= &  \sum_{c,d|c+d\,\,
{\rm odd}}
\!\!
F_{0,1}^{(\rm ground)}(\gamma\cdot\tau) 
\left\{\left[\frac{Z_\Gamma(\frac{3}{2}-\frac{3}{2}it-i\varphi)
Z_\Gamma(3-it-i\eta)}
{Z_\Gamma(\frac{5}{2}-\frac{1}{2}it-i\varphi)
Z_\Gamma(2-2it-i\eta)}
\right]_{\rm hol}\right\}_{\gamma} 
\nonumber \\
&\times &
\left\{\left[\frac{Z_\Gamma(\frac{3}{2}+\frac{3}{2}it-i\varphi)
Z_\Gamma(3+it-i\eta)}
{Z_\Gamma(\frac{5}{2}+\frac{1}{2}it-i\varphi)
Z_\Gamma(2+2it-i\eta)}\right]_{\rm antihol}\right\}_{\gamma},
\end{eqnarray}
where $\varphi \equiv \xi + \eta \equiv (\frac{1}{2}+\frac{d}{c})/{\rm Im}\,\tau$.
Let us consider the partition function with other spin structures, 
\begin{eqnarray}
\!\!\!\!\!
{\rm log}\prod_{m=2}^{\infty}|1-q^{m-\frac{1}{2}}|^2
& = & 
 \sum_{n=1}^{\infty}\frac{|q|^{-n}[q^{\frac{3n}{2}} +
\overline{q}^{\frac{3n}{2}}-|q|^{2n}(q^{\frac{n}{2}}+
\overline{q}^{\frac{n}{2}})]}{4n|\sin (n\pi \tau)|^2}
\nonumber \\
& = &
\sum_{n=1}^{\infty}\frac{e^{-n\pi {\rm Im}\,\tau}
\cos (3n\pi {\rm Re}\,\tau) -e^{-3n\pi {\rm Im}\,\tau}
\cos (n\pi {\rm Re}\,\tau)}
{2n |\sin (n\pi \tau)|^2}\,.
\end{eqnarray}
Then the contribution ${\widehat G}_{0,1}(\tau)$ becomes
\begin{eqnarray}
{\widehat G}_{0,1}(\tau) 
& = & 
\left[\frac{\prod_{n=2}^\infty (1-q^{n-1/2})} 
{\prod_{n=2}^\infty (1-q^{n})}\right]_{\rm hol}
\cdot
\left[\frac{\prod_{n=2}^\infty (1-\overline{q}^{n-1/2})}
{\prod_{n=2}^\infty (1-\overline{q}^{n})} \right]_{\rm antihol}
\nonumber \\
& = &
\left[\frac{Z_\Gamma(\frac{5}{2}-\frac{1}{2}it)
Z_\Gamma(3-it)}
{Z_\Gamma(\frac{3}{2}-\frac{3}{2}it)Z_\Gamma(2-2it)}
\right]_{\rm hol}
\left[\frac{Z_\Gamma(\frac{5}{2}+\frac{1}{2}it)
Z_\Gamma(3+it)}{Z_\Gamma(\frac{3}{2}+\frac{3}{2}it)
Z_\Gamma(2+2it)}\right]_{\rm antihol}\,. 
\label{G}
\end{eqnarray}
To this end we have
\begin{eqnarray}
G {\left[ \begin{array}{c} 0 \\ \frac{1}{2} \end{array} \right]}
(\tau) & = & 
\sum_{c,d|c+d\,\,
{\rm odd}} G_{0,1}^{(\rm ground)}(\gamma\cdot \tau)
{\widehat G}_{0,1}(\gamma \cdot\tau)
\nonumber \\
&= &  \sum_{c,d|c+d\,\,
{\rm odd}}
\!\! G_{0,1}^{(\rm ground)}(\gamma\cdot \tau) 
\left\{\left[\frac{Z_\Gamma(\frac{5}{2}-\frac{1}{2}it-i\varphi)
Z_\Gamma(3-it-i\eta)}
{Z_\Gamma(\frac{3}{2}-\frac{3}{2}it-i\varphi)
Z_\Gamma(2-2it-i\eta)}
\right]_{\rm hol}\right\}_{\gamma}
\nonumber \\
&\times &
\left\{\left[\frac{Z_\Gamma(\frac{5}{2}+\frac{1}{2}it-i\varphi)
Z_\Gamma(3+it-i\eta)}
{Z_\Gamma(\frac{3}{2}+\frac{3}{2}it-i\varphi)
Z_\Gamma(2+2it-i\eta)}\right]_{\rm antihol}\right\}_{\gamma}.
\end{eqnarray}

{\bf Quantum corrections to the black hole entropy.}
It has been claimed \cite{Maloney} that the formula for ${\bf Z}_ {0, 1}(\tau)$ is one-loop exact. Nevertheless it explicit calculation by evaluating the relevant one-loop determinants is not simple. One of the efforts to do so has been undertaken in \cite{Bytsenko}: The one-loop correction to the
Bekenstein-Hawking entropy for the non-spinning black hole 
was calculated by expressing the determinants in terms of the appropriate heat kernels. However, the final result obtained in \cite{Bytsenko} is not completely correct (in fact the authors 
used the cutt-off regularization procedure of the divergent volume of the fundamental domain, and as a consequence the structure of the group actions on a real hyperbolic space has been changed).
The purpose of this section is to get the correct contribution to the black hole entropy in terms of the Patterson-Selberg spectral function.

For our necessity remind that in (\ref{PF})
$24k= c_L=c_R= c$, and $c$ is the central charge of a conformal field theory. Then $q=\exp[2\pi i\tau]=\exp[2\pi(-{\rm Im}\tau +i{\rm Re}\tau)]$ such that 
$
{\bf Z}_{\rm cl}(\tau) = |q \overline{q}|^{-k}=
\exp [4\pi k {\rm Im}\tau]
$
corresponds to the classical prefactor of the partition function. 
The black hole partition function can be found by applying the modular transformation $\tau  \rightarrow -1/\tau$
to the partition function ${\bf Z}_{0,1}(\tau)$ of thermal $AdS_3$.
In fact the partition function (\ref{PF}) is a canonical ensemble partition function, so one may compute the black hole entropy using the following formula
\begin{eqnarray}
\!\!\!\!\!\!\!\!
S(\beta,\theta) & = & 
{\rm log}\,{\bf Z}_{1,0} - \beta {\bf Z}_{1,0}^{-1}
\frac{\partial \,{\bf Z}_{1,0}}{\partial \beta}
=
\left(1- \beta \frac{\partial}{\partial\beta}
\right){\rm log}~{\bf Z}_{1,0}
\nonumber \\
& = &
\left(1- \beta \frac{\partial}{\partial \beta}
\right){\rm log}
\left\{
{\bf Z}_{\rm cl}(\tau)\cdot \left[
\frac{Z_\Gamma(3- it)}
{Z_\Gamma(2- 2it)}\right]_{\rm hol}
\cdot\left[\frac{Z_\Gamma(3 + it)}
{Z_\Gamma(2 + 2it)}\right]_{\rm antihol}
\right\}
_{(\tau\rightarrow -1/\tau)}\,,
\label{S}
\end{eqnarray}
where $\beta \equiv {\rm Im}\,\tau = \ell/2\pi$ and
$t= {\rm Re}\,\tau/{\rm Im}\,\tau = \theta/2\pi\beta$.
In the limit $\theta \rightarrow 0$, i.e. for the non-spinning black hole, the entropy becomes
\begin{equation}
S(\beta, 0) = 
\left(1- \beta \frac{\partial}{\partial \beta}
\right){\rm log}
\left\{
{\bf Z}_{\rm cl}(\tau)\cdot \left[
\frac{Z_\Gamma(3)}
{Z_\Gamma(2)}\right]^2_{(\theta = 0)}
\right\}
_{(\tau\rightarrow -1/\tau)}\,.
\label{S}
\end{equation}

\subsection{Results on the holographic principle}
According to the holographic principle, there exist strong ties between certain field theories on a manifold ("bulk space")
and on its boundary (at infinity). A few mathematically exact results relevant to that program are the following. The class of Euclidean AdS$_3$ spaces (such as we have considered here) are quatients of the real hyperbolic space by a discrete group (a Schottky group). The boundary of these spaces can be compact oriented surfaces with conformal structure (compact complex algebraic curves). The results which can be regarded as manifestation of the holography principle are:
\begin{itemize}
\item{} There is a correspondence between spectral functions of hyperbolic three-geometry with it spectrum on a complex Riemann surface and Poincar\'e series associated with conformal structure in two dimensions. (The set of scatering poles coinsides with the zeros of a spectral function $Z_{\Gamma}(s)$; thus encoded in $Z_{\Gamma}(s)$ is the spectrum of a three-dimensional model).
\item{} An explicit correspondence exists between a certain class of fields in the bulk space (gravity) and the class of fields of conformal theory on the boundary.
\end{itemize}


\subsection*{Acknowledgments}
We are grateful to Professor L. Bonora for useful discussions.
The authors would like to thank the Conselho Nacional de Desenvolvimento Cient\'ifico e Tecnol\'ogico
(CNPq) for  support.

\end{document}